\documentclass[reqno]{amsart}
\usepackage[english]{babel}
\usepackage[latin1]{inputenc}
\usepackage{amsmath}
\usepackage{amsfonts}
\usepackage{amssymb}
\usepackage{times}
\numberwithin{equation}{section}
\newtheorem{theorem}{Theorem}[section]
\newtheorem{proposition}[theorem]{Proposition}
\newtheorem{lemma}[theorem]{Lemma}

\theoremstyle{definition}
\newtheorem{definition}{Definition}
\newtheorem{remark}[theorem]{Remark}
\newtheorem{remarks}[theorem]{Remarks}
\newcommand{\E}{\mathbb{E}}

\newcommand{\Z}{\mathbb{Z}}
\newcommand{\C}{\mathbb{C}}
\newcommand{\T}{\mathbb{T}}

\newcommand{\R}{\mathbb{R}}
\newcommand{\Schr}{Schr\"odinger }
\let\Im\undefined

\DeclareMathOperator{\Im}{Im}

\makeatletter
\newcounter{numcount}
\newcommand{\labelnummer}{\mbox{(\roman{numcount})}}%
\newenvironment{indentnummer}%
          {\let\curlabelspeicher\@currentlabel%
           \begin{list}{\labelnummer}{\usecounter{numcount}%
      \topsep1ex\partopsep2ex\parsep0pt\itemsep1ex\@plus1\p@%
                        \labelwidth2.5em\itemindent0em\labelsep1em%
                        \leftmargin3em}%
           \let\saveitem\item%
           \def\item{\saveitem%
                     \def\@currentlabel{\curlabelspeicher\labelnummer}%
                     \let\label\bemlabel}}%
         {\end{list}}%
\newenvironment{indentnummer*}%
          {\begin{list}{\labelnummer}{\usecounter{numcount}%
     \topsep1ex\partopsep2ex\parsep0pt\itemsep1ex
    \labelwidth2.5em\itemindent0em\labelsep1em%
         \leftmargin3.5em
                        }}%
         {\end{list}}%
\newenvironment{nummer}%
          {\let\curlabelspeicher\@currentlabel%
     \begin{list}{\labelnummer}{\usecounter{numcount}\leftmargin0em%
\topsep1ex\partopsep2ex\parsep0pt\itemsep0.5ex
         \labelwidth2.5em\itemindent3em\labelsep1em}%
           \let\saveitem\item%
           \def\item{\saveitem%
          \def\@currentlabel{\curlabelspeicher\labelnummer}%
                     \let\label\bemlabel}}%
         {\end{list}}%
\def\itemref#1{\expandafter\@setref\csname r@#1item\endcsname\@firstoftwo{#1}}%
\def\bemlabel#1{\@bsphack%
        \protected@write\@auxout{}%
               {\string\newlabel{#1}{{\@currentlabel}{\thepage}}}%
        \ifmmode\else%
        \protected@write\@auxout{}%
       {\string\newlabel{#1item}{{\labelnummer}{\thepage}}}%
        \fi%
        \@esphack}%
\makeatother
\begin{document}

\title[Quasi-Periodic Operators on Tree Graphs]{Persistence under Weak Disorder of AC Spectra of Quasi-Periodic  Schr\"odinger operators on Trees Graphs}

\author{Michael Aizenman}\author{Simone Warzel}
\address{Departments of Mathematics and Physics,
       Princeton University, Princeton NJ 08544, USA.}%

\email{aizenman@princeton.edu}\email{swarzel@princeton.edu}

\date{Submitted  (Mosc. Math. J.): 14 April  2005; Revised: 22 March 2006.}

\maketitle
\begin{abstract}
We consider radial tree extensions of one-dimensional quasi-periodic
Schr\"o\-dinger
operators and establish the stability
of their absolutely continuous spectra under weak but extensive
perturbations by a random potential.  The sufficiency criterion for
that is the existence of Bloch-Floquet states for the one dimensional
operator corresponding to the radial problem.\\[0.4cm]

{\it \normalsize Dedicated to Ya.~Sinai on the occasion of his seventieth birthday}\\[0.4cm]
\end{abstract}

\tableofcontents
\section{Introduction}

In this work we consider the effects of weak disorder on the
absolutely continuous (AC) spectra of tree extensions of
one-dimensional \Schr operators $H(\theta)$  with quasi-periodic (QP)
potential.
It is shown that contrary to the corresponding one-dimensional problem,
under certain conditions
AC spectra on trees persist under  perturbations by weak random potentials.

The operator $H(\theta)$  is of the form
\begin{equation}\label{eq:h}
	H(\theta) \, \psi_n := \psi_{n+1} + \psi_{n-1} + U(S^n
\theta)\, \psi_{n} \, ,
\end{equation}
where  $ U: \Xi \to \R $  is a continuous function on a
multidimensional torus $ \Xi \equiv [0,2 \pi)^\nu$
and $S$ is the shift
\begin{equation}
	 S \theta := (\theta + 2 \pi \alpha) \!\!\mod 2 \pi \, .
\end{equation}
with a frequency vector
$\alpha = (\alpha_1, \dots , \alpha_\nu) $ for which the action of
$S$ is ergodic.

To  $H(\theta)$  one may associate a ``fanned-out'' radial operator
${\widehat H}(\theta)$ which acts on the Hilbert space  $\ell^2(\T) $
over the vertex set $ \T $ of a
rooted regular tree graph as
\begin{equation}\label{eq:H}
	{\widehat H}(\theta) \, \psi_x := \sum_{y\in {\mathcal N}_x}
\psi_y + \sqrt{K} \, U\big(S^{|x|} \theta\big) \, \psi_x \, .
\end{equation}
In the above expression, the sum ranges over the set ${\mathcal N}_x$
of neighboring vertices of
$ x \in \T $,  the distance to the root is denoted by $ | x | $, and
$ K \geq 2 $ is the  branching number of the regular tree, i.e., the
number of forward neighbors of each vertex.
Next, disorder is added in the form of a random potential, which is
given
by a collection of independent and identically distributed
(iid) random
variables
$ \omega := \{ \omega_x\}_{x \in \T } $
associated with the vertices of $\T$.  In this fashion, one obtains
the ergodic operator
\begin{equation}
	{\widehat H}_\lambda(\theta,\omega) \, \psi_x = {\widehat
H}(\theta) \, \psi_x + \lambda \, \omega_x \, \psi_x
\end{equation}
where $ \lambda \in \R $ is the strength of the perturbation.

One dimensional Schr\"odinger operators can be studied by means of the associated cocycle, which in its projective representation is the skew product transformation
$(S, \mathcal{A}(E,\cdot))$ mapping the product space
$\Xi \times \C^+ $ as follows
\begin{equation}  \label{skew}
(\theta, \gamma) \mapsto (S \theta,
\mathcal{A}(E,\theta ) \gamma )
\end{equation}
where $ \mathcal{A}(E,\theta) $ is the M\"obius transformation
\begin{equation}\label{eq:Moeb}
	\mathcal{A}(E,\theta) \, :  \gamma \mapsto  \, U(\theta) -E
- \frac{1}{\gamma} \, \, .
\end{equation}

The issue of the stability of the AC spectra of ergodic radial
potentials on trees
was addressed in the recent work~\cite{ASW05}.  It was shown there
that AC spectra do not
disappear under weak disorder if for almost every $E\in \sigma_{\rm ac}(H(\theta))$
there is at most one (measurable) function
$\Gamma :  \Xi \to \C^+ $
whose graph is invariant under the action of the
Schr\"odinger cocycle, i.e., for which
\begin{equation}\label{eq:Schr}
	\mathcal{A}(E,\theta) \; \Gamma(\theta) = \Gamma(S\theta) \, .
\end{equation}

As is explained below, solutions of \eqref{eq:Schr}
in $ \C^+ $  are directly associated with
covariant eigenstates of  $ H(\theta) $.
For energies $E\in \sigma_{\rm ac}(H(\theta))$ one pair of conjugate
eigenstates is generally obtained through the Green function,
or alternatively the Weyl-Titchmarsh function (a generalization of
this function to trees is discussed in a related context in~\cite{ASW2}).

Here, we show that for operators with QP potentials the above
condition -- uniqueness
of solutions of \eqref{eq:Schr} -- is satisfied whenever $ H(\theta)
$  admits a
pair of Bloch-Floquet eigenstates with  Bloch momenta outside  of the countable
collection of resonant values corresponding to the spectrum of the
shift operator $ S $.  The existence of BF states at almost all energies in  the AC spectrum is well known to be related to the issue of reducibility of the \Schr cocycle to a constant~\cite{Eli92,PuigP,Puig05}.
Such reducibility plays also an essential role in our main
result.

The theory of Bloch-Floquet states for QP operators is a topic which
has been strongly developed
since the early works by Dinaburg and Sinai \cite{DiSi75}.
Their existence, which is required for our results,
was established by KAM and duality methods for a variety of
cases~\cite{DiSi75,BLT83,Sin87,CD89,Eli92,GJLS97,Jit99,BJ02,Puig05}, culminating in the recent proof by Avila and Krikorian~\cite{AvKr03} for which the sufficiency assumptions are just a Diophantine condition on $\alpha$ and smoothness ($C^\infty $) of the potential.

\section{Quasi-periodic Schr\"odinger operators on rooted tree graphs}
\subsection{Remarks on the AC spectrum}
Without the disorder, the AC spectrum of the fanned-out operator
   $ {\widehat H}(\theta) $  coincides with that of $ H(\theta) $, and
both can be
characterized in term of the  latter's Lyapunov exponent  $ \gamma(E) $
   (see  \cite{AvSi83} for its definition):
\begin{proposition}
For any QP potential and any $ \theta $
	\begin{equation}\label{eq:acspec}
	 {\sigma}_{\rm ac}\big( {\widehat H}(\theta) \big)  =
\sqrt{K} \, {\sigma}_{\rm ac}\big( H(\theta) \big)
	= \sqrt{K} \, \Sigma_{\rm ac}\, ,
	\end{equation}
	where $ \Sigma_{\rm ac} $ is the (Lebesgue) essential closure
of the set $
	\left\{ E \in \R \, : \gamma(E) = 0 \right\} $ .
\end{proposition}
\begin{proof}
It was noted in \cite[Prop.~A.1]{ASW05} that  $ {\sigma}_{\rm
ac}\big( {\widehat H}(\theta) \big)
	=  \sqrt{K} \, {\sigma}_{\rm ac}\big( H^+(\theta) \big)  $,
where $ H^+(\theta) $ is the restriction of \eqref{eq:h}
to the half line.   For the latter, Kotani theory and the statement
of independence from
$\theta$ proven in  \cite[Thm.~6.1]{LaSi99}, ensure that
$ {\sigma}_{\rm ac}\big( H^+(\theta) \big) = {\sigma}_{\rm
ac}\big( H(\theta) \big) = \Sigma_{\rm ac} $
for  all $ \theta $. \qed
\end{proof}
\begin{remarks}
	\begin{nummer}
	\item\label{i}
By  the Kotani principle,  the
presence of an AC component in the
spectrum of a one-dimensional
\Schr operator  requires its potential
to be
\emph{deterministic} \cite{Kot83,Sim83}.
For  QP potentials  this principle does not stand in
the way of $
H(\theta) $  having AC spectrum.  The existence of such
a component
was established  for various cases; an overview can be found
in~\cite{PF,Jit99,BJ02}.
\item  \label{rem:am}
The most studied QP operator is the \emph{almost Mathieu operator}
with the potential
$V_n(\theta) =  u\, \cos(2 \pi  \alpha n + \theta ) $, at an
irrational frequency $ \alpha \in (0,1)$.
In terms of \eqref{eq:h} it corresponds to $\nu =1$, i.e.,
$\Xi = [0,2\pi]$,  and
	\begin{equation}
	U(\theta) = u\, \cos(\theta) \;  .
	\end{equation}
As was shown in \cite{Jit99}
	for almost all $ \alpha $ and $ \theta $
	the spectrum of $ H(\theta) $ is
	\begin{itemize}
	\item pure absolutely continuous if $ u < 2 $,
	\item pure singular continuous if $ u = 2 $
	\item pure point with exponentially localized eigenfunctions
if $ u > 2 $.
	\end{itemize}
	\item\label{ex2}  A  broader class
of potentials, still with one frequency, is obtained by
letting  $ U(\theta) = u f(\theta) $ with a
real analytic, $2\pi $-periodic function $f$.  For the
corresponding QP \Schr operators  it was shown~\cite{BJ02} that for any
$ \alpha $ satisfying a  Diophantine non-resonance condition
$ \sigma\big(H(\theta)\big) = \sigma_{\rm ac}\big(H(\theta)\big) $ provided  $u$ is small enough.
\end{nummer}
\end{remarks}

The following observation may highlight the result presented below.
By the reduction of the radial spectral problem to the one dimensional
case, and the above mentioned Kotani principle \cite{Kot83,Sim83},
the AC spectra of the fanned-out operators are unstable under the addition
of arbitrarily weak radial disorder:
\begin{proposition}
If  $ \{ \omega_x \}_{x\in \T} $ are replaced by \emph{radial} iid random
variables, i.e.
   $ \omega_x = \omega_{|x|} $, for which $
\mathbb{E}\left[\log(1+|\omega_0| \right] < \infty $,
then the almost-sure AC spectrum vanishes,
   \begin{equation}
	\sigma_{\rm ac}\big( {\widehat H}_\lambda(\theta, \omega)
\big) = \emptyset ,
\end{equation}
for any  $ \lambda \neq 0 $.
\end{proposition}

The main result presented here  is that the response is different
when the disorder is  not constrained to be radial.

\subsection{A criterion for the stability of the AC spectrum}
The stability criterion which was derived in~\cite{ASW05} is expressed in terms
of the
uniqueness of solution of an equation associated with the \Schr cocycle
\eqref{eq:Schr}.  The equation can be viewed
as describing a covariant  \Schr  eigenstate, expressed in the
Ricatti form. Before explaining these concepts, let us present the
criterion in an explicit form.

\begin{proposition}[\cite{ASW05}]\label{prop:asw}
Let $ {\widehat H}_\lambda(\theta,\omega) = {\widehat H}(\theta) +
\lambda \omega $,
  with $ {\widehat H}(\theta)$ an operator of the form
\eqref{eq:H} and $\omega$   an iid random potential satisfying $
\E\left[\log(1+|\omega_0|)\right] < \infty $.
  If for Lebesgue-almost all $ E \in \Sigma_{\rm ac} $ the cocycle
equation associated with $ {\widehat H}(\theta) $:
\begin{equation}\label{eq:Schra}
	\Gamma(\theta) = \frac{1}{ U(\theta) - E - \Gamma(S\theta)}
\end{equation}
has not more than one measurable solution  for a function $ \Gamma:
\Xi \to \C^+ $
  with $ \Im \Gamma \geq 0 $,
then the almost-sure AC spectrum of $ {\widehat
H}_\lambda(\theta,\omega) $ is continuous
at $ \lambda = 0 $, in the sense that for every Borel subset $I
\subseteq \Sigma_{\rm ac} $
	\begin{equation}\label{eq:stab}
		\lim_{\lambda \to 0} \left| \, \sigma_{\rm
ac}\big({\widehat H}_\lambda(\theta,\omega)\big) \, \cap \, \sqrt{K}
I \, \right| =
		\sqrt{K} \left| I \right| \, ,
	\end{equation}
where $ | \cdot | $ denotes Lebesgue measure.
\end{proposition}
\begin{remark} The result proven in~\cite{ASW05} includes also the statement
that in a suitable sense the AC density of the spectral measure associated with the
root vector is $L^1$-continuous at $\lambda=0$.
Moreover, the derivation used a weaker condition than independence
  for the disorder variables $\omega$.  Rather, it suffices there to
assume only weak correlations for  the joint distributions along
any two disjoint forward subtrees of $\T$.   The results presented
here are directly applicable also to this generalization.
\end{remark}

To shed some light on \eqref{eq:Schra}, which is equivalent to
\eqref{eq:Schr},
let us note that  the Schr\"o\-dinger equation
$ \phi_{n-1} + \phi_{n+1} + (V_n - E) \, \phi_n = 0 $,  can
equivalently be expressed in
terms of the Ricatti variables
$\gamma_n := - \phi_n / \phi_{n-1}$ as:
\begin{equation} \label{eq:gamma}
\gamma_n  \ = \  \frac{1}{V_n - E - \gamma_{n+1} } \, .
\end{equation}

If the eigenstate state $\phi$ can be chosen as a
covariant function of $\theta$, then
$\Gamma :=\gamma_0[\phi]$ will satisfy \eqref{eq:Schra}.   By  a
{\em covariant eigenstate} we mean here a generalized
eigenfunction whose $\theta$-dependence is such that
\begin{nummer}
    \item $ H(\theta) \, \phi(\theta) \ = \ E\ \phi(\theta)  $,
    \item $ \phi_{n+1}(\theta) = e^{i \kappa(\theta)}  \phi_n(S\theta)
\; $ for some measurable
$  \kappa: \Xi \to \R $,
\end{nummer}
where (i) is to be taken in the weak sense, as appropriate
for a generalized eigenfunction.

As an aside, let us note that the converse is also true.
More
precisely, if $ \Gamma: \Xi \to \C^+ $ is a solution of
\eqref{eq:Schra}
with $ \Im \Gamma > 0 $ then
\begin{equation}\label{eq:DeSi}
	\phi_{-1}(\theta) := \frac{1}{\sqrt{\Im \Gamma(\theta)}} ,
\qquad \phi_0(\theta) := \phi_{-1}(\theta) \, \Gamma(\theta)
\end{equation}
determine a covariant eigenstate of $ H(\theta) $,
as was noted already  in~\cite{DeSi83,Kot83}.

The  difference between a covariant eigenstate and a covariant
eigenfunction is significant.  States are interpreted as rays,
for which the phase factor in~(ii) has no
effect.   Covariant eigenfunctions, interpreted as solutions of~(i)
with $\kappa =0$ in~(ii) may not exist, or exist only at isolated
energies.
However, solutions
with  non-zero, constant $\kappa $ do occur, and form the appropriate
generalization of the Bloch-Floquet states to QP operators.
We present their definition in the next section, where we also
state our main result.

\begin{remark}

For the analysis of~\cite{ASW05} it was relevant
that the diagonal element of the Green function of $ H^+(\theta) $ taken at the root $ 0 $,
\begin{equation}\label{eq:Green}
	\Gamma(E,\theta) = \big( H^+(\theta) - E - i0\big)^{-1}(0,0) \, ,
\end{equation}
is a solution of \eqref{eq:Schra} with $ \Im \Gamma(E,\theta) > 0 $
for almost every $ E \in
\sigma_{\rm ac}(H(\theta)) $.
The proof of Proposition~\ref{prop:asw}
proceeds by  establishing that in any limit $\lambda, \eta \to 0$
the distribution of the diagonal Green function
$ \widehat\Gamma_\lambda(E+i\eta ,\theta,\omega) $
of $ {\widehat H}_\lambda(\theta, \omega) $
is of vanishing width, in the dependence on $\omega$ ~\cite{ASW05}.
The subtle point is that the values may still depend on the way
$\lambda$ and $ \eta $ approach their limit.  Any accumulation point
satisfies the cocycle equation \eqref{eq:Schra}, but that in itself
does not yet allow to conclude that it coincides with
$K^{-1/2}\, \Gamma(E/\sqrt{K},\theta)  $.
However, the  uniqueness of solution  which is discussed here
guarantees the distributional convergence of
$ \widehat\Gamma_\lambda(E+i\eta ,\theta,\omega) $,
and by implication the continuity of the AC spectum.
\end{remark}

\section{Bloch-Floquet eigenfunctions and the main result}
\begin{definition}\label{def:Bloch}
A \emph{Bloch-Floquet (BF) eigenfunction} of $ H $ with energy $ E\in
\R $ and quasi-momentum $ k \in(-\pi , \pi]  $ is a non-vanishing
function
$ \psi: \Z \times \Xi \to \C $ with the properties:
\begin{nummer}
\item $ H(\theta) \, \psi(\theta)\ = \ E \, \psi(\theta) \qquad $
(in the weak sense).
\item $ \psi_n(\theta) = e^{ikn}\,\varphi(S^n\theta) \quad$ for some
continuous $\varphi: \Xi \to \C $.\\[-2ex]
\end{nummer}
We say that  $ \psi $ forms part of a \emph{conjugate BF
pair} iff  $ \psi(\theta) $ and
$ \overline{\psi(\theta)} $ are linearly independent for almost every
$ \theta $.
\end{definition}
\begin{remarks}
\begin{nummer}
	\item The second requirement in the above definition is
equivalent to the covariance property
	\begin{equation}
		 \psi_{n+1}(\theta)= e^{ik} \, \psi_n(S\theta) \, .
	\end{equation}
	\item\label{rem:G}
	Bloch-Floquet eigenfunctions come naturally in pairs.
	Namely, if $ \psi $ is one then its complex-conjugate $
\overline{\psi}$
	is also
	a BF eigenfunction with energy $ E $ and reversed quasi-momentum.
	Their Wronskian
	\begin{equation}\label{eq:Wronskian}
	  \big[\, \psi,\overline{\psi}\, \big](\theta) :=
	\psi_0(\theta) \, \overline{\psi_{-1}(\theta)} -
\overline{\psi_0(\theta)} \,\psi_{-1}(\theta) \, ,
	\end{equation}
	is
	independent of $ \theta $. Therefore, if $ \psi(\theta) $ and
$ \overline{\psi(\theta)} $ are linearly independent
	for some $ \theta $, they are linearly independent for all $
\theta $ and
	$ \big[\, \psi,\overline{\psi}\, \big] \neq 0 $.
	This implies that the ratio
	\begin{equation}\label{def:G}
		\gamma(\theta) := -\frac{\psi_0(\theta)}{\psi_{-1}(\theta)}
	\end{equation}
	is well-defined and takes values with either $ \Im
\gamma(\theta) > 0  $ or $  \Im \gamma(\theta) < 0 $ for all $ \theta
$.
\end{nummer}
\end{remarks}

Our main observation, whose proof is presented below, is that the
existence of a conjugate BF pair of $ H $
at almost all energies in $I$ allows one to conclude that the cocycle equation~\eqref{eq:Schra}
has a unique solution and hence
the criterion of Proposition~\ref{prop:asw} is met.  That yields:

\begin{theorem}\label{thm:main}
	In the situation of Proposition~\ref{prop:asw} let $  I
\subseteq \Sigma_{\rm ac} $ be a Borel set such that at
	Lebesgue-almost all $ E \in I $, $ H $
	admits a conjugate pair of Bloch-Floquet states.
	Then \eqref{eq:stab} holds.
\end{theorem}

\section{Uniqueness of the covariant solution of the projective
\Schr equation}

\subsection{BF states and reducibility of the \Schr cocycle}

More frequently, the discussion of the \Schr cocycle is carried out in the space $\Xi \times  SL(2,\R) $.  The role  of
$\mathcal{A} (E,\theta)$ in
\eqref{skew} is taken by the matrix
\begin{equation}\label{eq:defA}
	A(E,\theta) := \left( \begin{matrix} E - U(\theta) & \, -1 \\
						1 &\; 0 \end{matrix}
\right) \, .
\end{equation}
In this notation, the existence of BF pairs is known to be equivalent to the \emph{reducibility} of this cocycle to a constant one:

\begin{proposition}[cf.~\cite{Puig05}]\label{prop:red}
The following two statements are equivalent:
\begin{indentnummer}
\item $ H $ admits a conjugate BF pair with energy $ E $ and
quasi-momenta $ \pm k $.
\item\label{prop:red1} the Schr\"odinger cocycle is reducible to a constant
	matrix,
	\begin{equation}\label{eq:red1}
	Z(S\theta)^{-1} \,  A(E,\theta) \, Z(\theta) \, =  \, \left(\begin{matrix}
e^{-ik} &  0 \\ 0 &  e^{ik} \end{matrix}\right)
	\end{equation}
	where  $ Z(\theta) = : \left(\begin{matrix}
\overline{\psi_0(\theta)} &  \psi_0(\theta) \\
					\overline{\psi_{-1}(\theta)}
& \psi_{-1}(\theta) \end{matrix}\right) $ is not
	singular, i.e.,  $ \det Z(\theta) \neq 0 $.
\end{indentnummer}
\end{proposition}
\begin{remark}
If $ H $ admits a conjugate pair of covariant eigen\emph{states},
which according to~\eqref{eq:DeSi}
holds for almost every $ E \in \sigma_{\rm ac}(H(\theta)) $, then
\eqref{eq:red1} remains true if $ k $ is replaced by $ \kappa(\theta)
$.
\end{remark}

Historically, reducibility was first established
for  various examples of QP operators $ H $ for almost all energies
in their AC spectra
\cite{DiSi75,BLT83,Sin87,CD89,Eli92,Puig04,GJLS97,Jit99,BJ02,Puig05}.
In particular, it was proven to hold for the almost Mathieu operator and the
family of real analytic QP potentials mentioned in
Remarks~\ref{rem:am} and ~\ref{ex2}.
In this context, it was natural to ask
whether all quasi-periodic  operators admit conjugate
BF pairs for almost every $ E \in \Sigma_{\rm ac} $.
This question was raised in \cite{DeSi83}, and
was disproved in \cite{Las93} in the general case.
However, for shifts $ S $ on the one-dimensional torus with certain Diophantine frequencies $ \alpha $ and arbitrarily often differentiable $ U $, it was
shown in \cite{AvKr03} that BF pairs indeed exist for almost every $ E \in \Sigma_{\rm ac} $.

\subsection{Reducibility and uniqueness}

The main observation leading to Theorem~\ref{thm:main} is:
\begin{lemma} \label{lem:unique}
Assume that $ H $ admits a conjugate BF pair with energy $ E $
and quasi-momenta $ \pm k $ with $ | k |/ \pi  \,  \notin  \left\{ \big( m \cdot \alpha
\big)\!\! \mod \Z \, : \, m \in \Z^{\nu} \right\} $. Then
\eqref{eq:Schr}
	has a unique solution with values in $ \C^+ $.
\end{lemma}
\begin{remark}\label{rem:ids}
The above condition on the quasi-momenta can be rephrased in terms of the integrated density of states (IDS) $ n(E) $
of $ H(\theta) $ (see \cite{AvSi83} for a definition) using its relation to the rotation
number of the \Schr cocycle \cite{DeSo83}. Namely, there exists some $ m \in \Z^\nu $ such that
	\begin{equation}
	 \frac{| k |}{2\pi} = \Big( \,\frac{n(E)}{2} +  m \cdot
\alpha \, \Big) \!\! \mod \Z \, .
	\end{equation}

\end{remark}
\begin{proof}[Proof of Lemma~\ref{lem:unique}]
The M\"obius transformation $ \mathcal{A}(E,\theta)  $ given by
\eqref{eq:Moeb} is just a projective counterpart of
$ A(E,\theta) $ in \eqref{eq:defA}. From Proposition~\ref{prop:red} it thus follows that $\mathcal{A}(E,\theta) $ is
reducible to a constant cocycle, in the sense that there exists a M\"obius
mapping $\mathcal{Z}(\theta)$ such that
	\begin{equation}\label{eq:red2}
	\mathcal{A}(E,\theta) \, \mathcal{Z}(\theta) = \mathcal{Z}(S
\theta) \, e^{-2 ik} \, .
	\end{equation}
An explicit expression for $\mathcal{Z}(\theta)$, in terms of the BF
state $\psi(\theta)$, is
	\begin{equation}
	 	\mathcal{Z}(\theta)\, : \gamma \, \mapsto \, \gamma =
		\frac{\overline{\psi_0(\theta)}\, \gamma -
\psi_0(\theta)}{- \overline{\psi_{-1}(\theta)} \, \gamma +
	\psi_{-1}(\theta)} \, .
	\end{equation}
As was noted before, $ \gamma(\theta) := -\psi_0(\theta) /
\psi_{-1}(\theta) $ satisfies \eqref{eq:Schr} and
	according to Remark~\ref{rem:G}
	we may assume without loss of generality
	that   $ \Im \gamma(\theta) > 0 $ for all $ \theta $.
	Suppose now there exists yet another solution $ \widetilde\gamma \neq \gamma $  of
	\eqref{eq:Schr} with values in $ \C^+ $. Then
	\begin{equation}
	f(\theta) := \mathcal{Z}(\theta)^{-1}
\widetilde\gamma(\theta) =
\frac{\psi_{-1}(\theta)}{\,\overline{\psi_{-1}(\theta)}\,} \,
	\frac{\widetilde\gamma(\theta) - \gamma(\theta)
}{\widetilde\gamma(\theta) - \overline{\gamma(\theta) }\,}
	\end{equation}
	satisfies
	\begin{equation}\label{eq:eigenvalue}
		 \big(\mathcal{S} f\big)(\theta) := 	f(S \theta) =
\exp\left[ -2i k  \right] \, f(\theta)
	\end{equation}
	where we have introduce the unitary Koopmann operator
	$ \mathcal{S} : L^2(\Xi)  \to L^2(\Xi) $
	associated with the ergodic shift $ S $.
	For a QP shift with frequency $ \alpha $
		its spectrum consists of the countable set
	\begin{equation}
   	{\rm spec}(\mathcal{S}) = \left\{ \, \exp( 2 \pi i \, m \cdot
\alpha ) \, : \, m \in \Z^{\nu} \right\} \, .
	\end{equation}
	According to \eqref{eq:eigenvalue}, $ f $ is a proper eigenfunction
	of the Koopmann operator $ \mathcal{S} $. Since $ \exp\left[
\pm 2i |k| \right] \notin {\rm spec}(\mathcal{S}) $ this implies that
	$ f = 0 $, which is a contradiction.
\qed
\end{proof}

\section{Proof of the main result}

We are now ready to complete the prove of our main result.

\begin{proof}[Proof of Theorem~\ref{thm:main}]
In view of Proposition~\ref{prop:asw}, Lemma~\ref{lem:unique} and Remark~\ref{rem:ids}
it remains to show that for almost all $ E \in \Sigma_{\rm ac} $
\begin{equation}\label{eq:nspec}
	n(E)  \notin \left\{ \big( m \cdot \alpha \big)\!\! \mod \Z
\, : \, m \in \Z^{\nu} \right\} \, .
\end{equation}
But this follows from the fact that $ \Sigma_{\rm ac} $ is a set of
positive Lebesgue measure on which the integrated density
of states is not constant \cite{AvSi83}. Since the right side in
\eqref{eq:nspec} is a countable set, it cannot coincide with
the image of $ \Sigma_{\rm ac} $ under the map $ E \mapsto n(E)  $. \qed
\end{proof}
\begin{remark} By the gap-labeling theorem \cite{JoMo82,DeSo83} the
condition $n(E)  = \big( m \cdot \alpha \big)\!\! \mod \Z  $, for some
$ m \in \Z^{\nu} $,
characterizes the spectral gaps
of $ H(\theta) $.
\end{remark}

\section*{Acknowledgement}

We thank Robert Sims, Svetlana Jitomirskaya, Michael Goldstein and
Uzy Smilansky for stimulating discussions of topics related to this
work.
MA thanks for the gracious hospitality enjoyed at the Weizmann Institute.
This work was supported in parts by the Einstein Center for
Theoretical Physics and the Minerva Center for Nonlinear Physics at
the Weizmann Institute, by the US National Science Foundation, and by
the Deutsche Forschungsgemeinschaft.

\end{document}